# Thermally expanded graphite polyetherimide composite with superior electrical and thermal conductivity

Fatema Tarannum, Swapneel Danayat, Avinash Nayal, Rajmohan Muthaiah, Roshan Sameer Annam, Jivtesh Garg

School of Aerospace and Mechanical Engineering, University of Oklahoma, Norman,73019, USA

**Abstract:**

Thermally expanded graphite (EG) polymer composite has high potential in modern thermo-electronic industry due to cost effectiveness, excellent electrical and thermal properties. In this study, we report that expanded graphite polyetherimide (EG/PEID) composite has been fabricated by simple solution mixing technique, reinforcing a low cost EG filler into host polymer matrix, polyetherimide (PEID). This technique enables the inclusion of 3D graphitic filler into polymer matrix retaining the worm structure of expanded graphite, thus developing continuous network throughout the composite. In result, ~19 orders of magnitude rise of the polymers' electrical conductivity up to 969 S/m has been achieved upon the inclusion of the 10wt% of EG filler. Theoretical prediction using effective medium approach reveals the impact of an imperfectly conducting interface and interfacial tunneling on electrical conductivity of the composites. We also report that 10wt% EG filler composition composite exhibits the thermal conductivity ($k$) value of 7.3W/mK, implying significant enhancement of 3070% compared to pure PEID. Scanning Electron Microscopy (SEM) represents the morphology of fillers and composites, revealing the preserved structure of EG to develop the conductive pathway between polymer and filler. Investigation on the structural properties before and after the thermal expansion of EG fillers has been performed using Raman, X-ray diffraction (XRD) and X-ray Photoelectron Spectroscopy (XPS) analysis. In summary, a simple fabrication route of solution mixing has been demonstrated to achieve an outstanding electrically and thermally conductive EG/PEID composite without compromising the efficiency of EG filler structure. The expanded graphite-based polymer composites can retain their outstanding properties in wide range of thermal and electronic industry.

# 1. Introduction

Carbon based fillers has intrinsic electrical properties, comprising excellent mobility of charge carrier, π electrons with low-energy band, and outstanding optical transparency perpendicular to the transmitted light[1]. However, carbon-based fillers can't directly be implied into electronic devices, requiring host material to be manufactured with adaptive physical properties[2]. Polymer materials offer flexible physical properties and excellent processability but has lacking in free charge transportation. Also, polymer possess very poor thermal conductivity (<0.5 W/mK)[3]. To manufacture desirable products with superior electrical and thermal performance, carbon-based fillers[4-11] have been added to the polymer matrix. These carbon fillers (such as graphene, CNT) based polymeric composites has wide range of applications including sensors[12, 13], electromagnetic interference shielding[14, 15], batteries[16, 17], supercapacitors[18, 19] and functional materials[20, 21].

Electrical conductivity (EC) of composites has been largely influenced by the composite preparation methods, embedded polymer matrix, dispersion quality and size and amount of filler[22]. It is critically important to make a continuous network path in the composite to achieve higher conductive property. Graphene nanosheet (2D) or carbon nanotube (CNT, 1D) with rod or disk like shape, can enable continuous conductive network in composites due to their high surface area and aspect ratio[23-25]. But those require higher content of filler to achieve percolation threshold, raising difficulties in embedding filler into matrix. 3D continuous network from 2D filler using chemical vapor deposition[26-29] or solution-based synthesis[30] has been developed for outstanding electrical and thermal properties of composite. Kashi et al.[31] reported an electrical percolative 3D network of GNP/polylactide (PLA) nanocomposite at 6–9 wt% GNP filler composition. In another study, Moghaddam et al[32] developed interconnected 3D network within graphene oxide (GO)/epoxy composite using functionalized GO filler, indicating the superiority in electrical conductivity due to successful 3D network formation. Such 3D network of filler can be utilized for continuous electron transfer in composite, thus overcoming the barrier impending from polymer medium, leads to efficient electrical conduction mechanism.

Thermally expanded graphite (EG) has been a promising 3D filler with lightweight, porous structure (pore size varies from macro to nanoscale) to manufacture composites for wide range of application because of its excellent electrical conductivity ($10^6$–$10^8$ $Sm^{-1}$) and thermal conductivity (>300 $Wm^{-1}K^{-1}$). EG can be easily obtained from the synthesized graphite, in another word, graphite intercalated compound (GIC) or expandable graphite (EPG) through thermal treatment. In this work, we have used thermally expanded graphite and fabricated EG/polymer composite via solution casting technique. EC value of 969 S/m at very low concentration of 10wt%

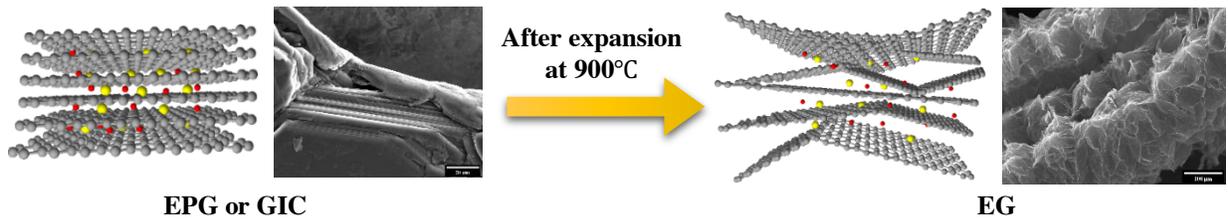

EPG or GIC        After expansion at 900°C        EG

Figure 1 a) Schematic preparation of expanded graphite from expandable graphite (EPG) at 900°C with FESEM images of EPG, and EG

EG filler content has been reported in this study. After uneven thermal expansion, 3D hierarchical structure of EG principally covalently bonded with 2D graphene nanosheets over the length scale (as shown in Fig 1), imparts superior electrical and thermal performance in composite.

Nevertheless, challenges remain with the feasibility of preparation method and amount of EG filler content used for the polymer composites to gain excellent electrical and thermal properties. Melt mixing[33-35], high shear mixing[36, 37], ball milling technique[38, 39] have been applied to fabricate the EG/polymer composites. High shear force during melt mixing technique was not found suitable for several studies due to the poor dispersion and segregation issue[40, 41]. Due to fragile nature of expanded graphite, use of EG filler is challenging to disperse into polymer without any damage. Manufacturing composite via melt mixing or compression molding technique requires higher filler concentration during preparation of composite. According to Wei et al[42], 13.33vol% expanded graphite filler, pre-melt blended with stearic acid and polyethylene wax was melt-mixed with low density polyethylene (LDPE) to obtain EC of 1345 S/m. Yang et al.[43] fabricated multilayer plastic packaging waste (MPW)/EG composite using powder mixing technique followed by compression molding and achieved EC of 253 S/m for 31.6vol% EG filler.

Significant research has developed polymer composites using EG fillers to obtain EC at lower percolation threshold. She et al. embedded expanded graphite into a high-density polyethylene (HDPE) matrix via melt-extrusion and found a percolation threshold of ~5.7wt% for HDPE/EG conducting composite[44]. Thermally expanded graphene oxide polycarbonate nanocomposites exhibited percolation at 1.25wt % for surface resistivity and reached upto 2×10$^5$Ω surface resistivity at 3wt% for nanocomposites[45]. Piao and his co-workers developed a EG polymer composites using a conducting polymer poly(3,4-ethylenedioxythiophene) poly-(styrene sulfonate) (PEDOT:PSS) and that yields EC up to 10$^4$ S/m for 50wt% EG filler[46]. The graphene–polyimide nanocomposites exhibited an ultra-low graphene percolation threshold of 0.03 vol % and maximum dc conductivity of 0.94 S/cm[1]. Li et al. prepared poly(vinylidene fluoride)PVDF/EG composites and found percolation threshold of 6.3 vol% for electrical conductivity[47]. Several studies also report the *k* value of EG/polymer composite. Sari et al. [48] found that 10 wt% EG the thermal conductivity of the paraffin/EG composite material was increased to 0.82 W/mK at 10wt% filler due to proper absorbance of paraffin into expanded graphite. Inclusion of EG filler into polyethylene glycol (PEG) improves ~4.4-fold higher thermal conductivity compared to pure polyethylene glycol conductivity of 0.2985Wm$^{-1}$K$^{-1}$ [33]. Kim et al[49] achieved 10.77 W/mK for EG/epoxy composite at 20wt% inductively coupled plasma treated EG filler amount using ultrasonication for 30min. Incorporating lower amount of filler to achieve higher electrical and thermal performance has been presented in this study along with detail morphological analysis of the EG filler and composite.

Polar polymer, polyetherimide (PEID) chosen as polymer matrix in this work, is a high-performance polymer with excellent flame retardancy, chemical resistance and high temperature stability[50]. High melting temperature of PEID, 340–360 °C makes the fabrication process challenging[51]. Organic solvent such as dimethylacetamide has potential in processing this polymer at low temperature of 130°C, which makes the solution casting technique suitable and straightforward to mix with EG filler. Furthermore, this strategy allows the homogeneous solution of PEID (in solvent) to fill the pores of EG with polymer and preserve the porous 3D structure of EG in composite. Particularly, solution casting technique takes the advantage of minimal distortion avoiding any shear force during sample making. Additionally short time ultrasonication disperses the EG filler well into polymer matrix. Oxygen groups of EG interacts with imide groups of PEID through hydrogen bonding (as shown in fig 2b), enables better interaction between polymer matrix and filler. X-ray photoelectron spectroscopy (XPS) analysis

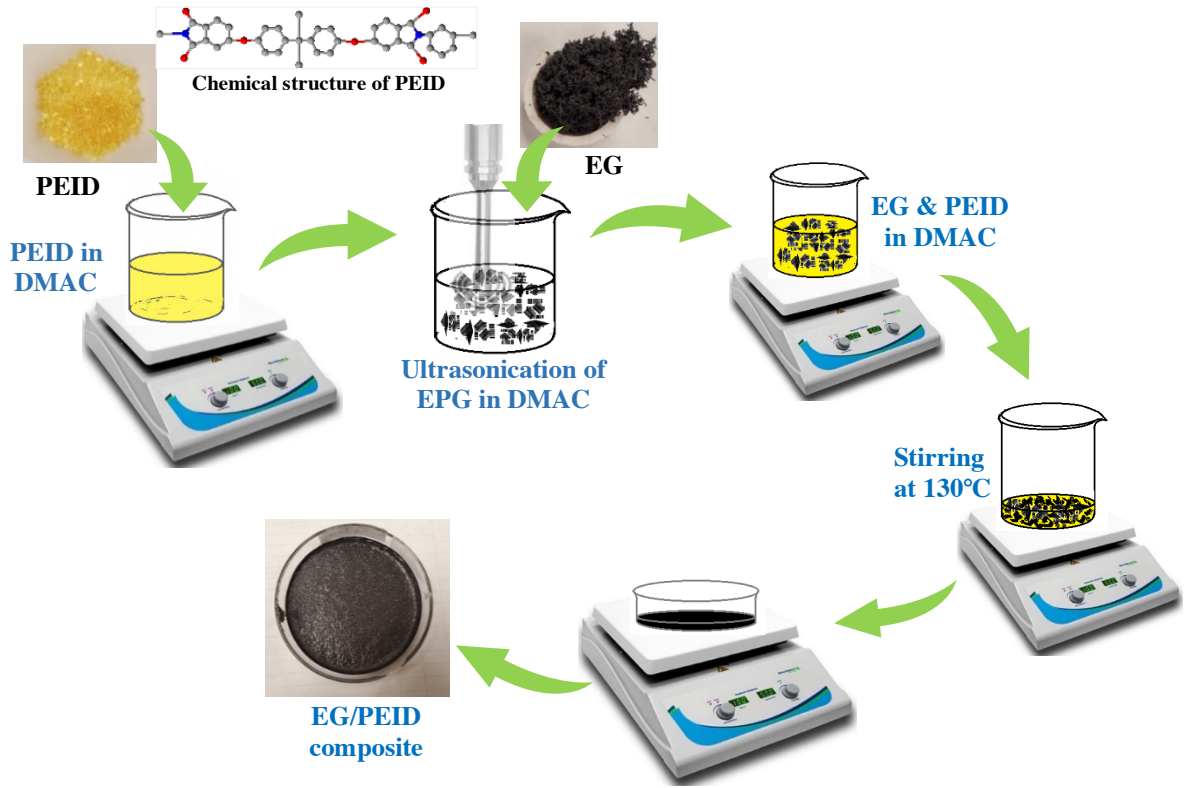

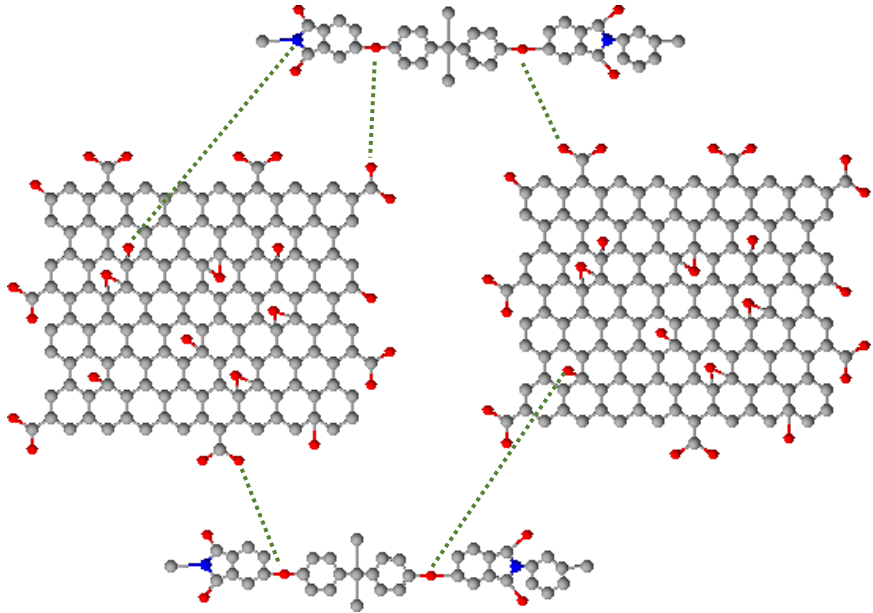

Figure 2a) Schematic preparation of EG/PEID composite, b) Interaction between polyetherimide (PEID) and graphite nanosheet through hydrogen bonding

provides the evidence for the presence of oxygen content on EG filler.

In this paper, we demonstrate an effective solution mixing route with short time ultrasonication for fabricating EG/polyetherimide (PEID) composite to mobilize the insulating characteristic of polyetherimide ($10^{-17}$S/m) upto maximum in-plane electrical conductivity of 969 S/m (above the percolation threshold of 1wt%). The insulative polymer starts to change its behavior in electron movement as EG filler is added to polymer. Percolation threshold at specific filler loading turns the insulating behavior of polymer into conductive through network formation in the composites. Percolative behavior is studied with significant research studies regarding electrical conductivity of expanded graphite polymer composite. Instead of direct contact, connection through tunneling mechanism takes place in this percolative region, forming an interface between the filler and the matrix[52]. Zare et al. developed a model to show the effect of interphase and tunneling distances on electrical conductivity of graphene/polymer composites. It revealed that high number of contacts between graphene nanosheets diminish the tunneling distance and interfacial resistance between adjacent nanosheets, resulting into better conductive network[53]. For EG/PEID composite, a continuous 3D network of EG filler establishes a percolative network in composite at lower content of filler, leading to higher EC value of composite at 10wt% filler composition. As large graphite nanosheets are covalently bonded over the length of EG, causing reduction in tunneling distance as well as increase in number of contacts between graphite nanosheets, thus results in lower interfacial resistance between graphite nanosheets. We have explored this mechanism on EG/PEID composite through the effective medium theory including filler loading, percolation threshold, conductivity effected by interfacial resistance, and electron tunneling. We compared the EC value of EG filler-based composite to the graphene nanoplatelet polymer composite.

In addition, we have investigated thermal properties of the solution casted EG/PEID composites. Thermal conductivity of the composite reaches upto 7.3 W/mK, 32-fold higher over pure polyetherimide (0.23 W/mK). The dispersion and morphology of the expanded graphite/PEID composite have been investigated using Field Emission Environment Scanning Electron Microscopy (FE-ESEM). The presence of functional groups, structural integrity and defective state of expanded graphite and graphite intercalated compound were characterized by XPS, XRD and Raman analysis to understand the changes occurred after expansion.

## 2. Experimental Section

### 2.1. Materials

Expandable graphite (EPG) or graphite intercalated compound (GIC) or, with an average particle size of ≥300 μm ( +50mesh size: ~92%) and expansion ratio 280:1 (ASB-3772) were bought from Graphite store[54]. Graphene Nanoplatelets of 15 nm thickness and lateral size of ~25 μm were bought from Sigma Aldrich[55]. Also, polyetherimide (PEID) pellets (melt index of 18 g/10 min and a density of 1.27 g/mL) and N, N-dimethylacetamide (DMAC) were purchased from Sigma Aldric[55] and Alfa Aesar[56] respectively.

### 2.2. Fabrication of the EG/PEID composites

Thermal expansion of expandable graphite was carried out to obtain the worm structured expanded graphite (EG) using hot furnace at 900°C. The expandable graphite was kept inside the furnace for very short time, ~1min to achieve the highest expansion. Solution casting technique was used to prepare the expanded graphite polymer composite and polyetherimide pellets was chosen as polymer matrix to mix with expanded graphite filler. Fig 2 represents the fabrication procedure of the EG/polyetherimide composite. EG fillers so obtained were then dispersed into 20 mL DMAC. Separately PEI pellets were dissolved using 50 mL DMAC at 130 °C for 1 h. Both the DMAC solution with EG filler and dissolved polymer were mixed together and blended for 3 hr at 130 °C followed by ultrasonication at 20% amplitude for ~ 120 sec. Then, the mixture was cast into a glass petri dish and kept at 100 °C. The composite film was peeled off after 24-48 hr. The composite films were prepared for different concentration of filler- 2.5, 5, 7.5 and 10 wt% EG/PEID composite. Different ultra-sonication time (30sec -150sec) have been used to optimize the preparation process. To compare the electrical property with graphene-based polymer composite, the composite films were also prepared using graphene nanoplatelet following the same procedure.

## 3. Characterization:

### 3.1 Measurement of Electrical Conductivity:

In-plane EC of EG/PEID and GnP/PEID composites was measured with using a four-point probe setup at room temperature. Keithley Multimeter 2100 (Keithley Instruments, Solon, OH, USA) coupled with in house built four probe point set up was arranged to measure the resistance of composite film. EG/PEID films were cut into pieces with length = 4.3mm, width = 1cm and thickness = 0.3-0.5 µm to determine the average resistivity using the resistance,($\Omega$) and dimensions of the flat strips. Volume resistivity was calculated using following equation,

$$\rho = \frac{R \times A}{L} = \frac{R \times w \times t}{L}$$

Where, R, w, t and L represent the resistance ($\Omega$), width, thickness and length of the sample respectively. Electrical conductivity is inverse of resistivity, calculated by $\sigma = \frac{1}{\rho}$.

### 3.2 Thermal Conductivity (k):
k of EG/polymer composites was measured by the laser flash technique. A Netzsch LFA 467 Hyperflash (Netzsch, Germany) laser was used to measure the through-thickness thermal diffusivity of the samples. 8-12 samples of 12.5mm diameter and 0.3-0.4 mm thickness were used to measure the thermal diffusivity ($\alpha$) at room temperature (23 °C). The samples were coated with a thin layer of graphite paint before the measurement to efficiently absorb heat from a flash lamp, and average of 6-8 measurement is reported. This laser flash technique induces heat by a laser pulse on one surface of the sample and the temperature rise is captured on the other surface of sample as a function of time. $\alpha$ is determined by LFA using the following equation: $\alpha = (0.1388\, d^2)/t_{1/2}$, where, $\alpha$ is the thermal diffusivity (mm$^2$/s), $t_{1/2}$ is the time to obtain half of the maximum temperature on the rear surface, and d denotes the sample thickness (mm). The thermal conductivity was calculated using k = $\alpha\rho$Cp, where k, $\rho$, and Cp represent the thermal conductivity, density, and specific heat constant of the sample, respectively. In this work, density and specific heat of the composite samples were calculated using gas pycnometer (AccuPyc II 1340, Micromeritics Instrument Corporation, USA) and Netzsch differential scanning calorimetry (DSC) (DSC 204F1 Phoenix, Netzsch, USA).

**3.3 Scanning Electron Microscopy (SEM):** Morphological characterization of EG filler and EG/polymer composites was carried out by high resolution Field Emission Environmental Scanning Microscopy (Quattro S FE-ESEM, Thermofisher Scientific, USA). This SEM was operated in secondary electron (SE) mode at an accelerating voltage of 20 kV.

**3.4 Raman Spectroscopy (RS):** Raman spectroscopy (RS) was performed using a DXR3 SmartRaman Spectrometer (Thermofisher Scientific, USA) to collect the data over the range from 3000 to 1000 cm$^{-1}$, laser wavelength $\lambda_L$ = 633 nm, spectral resolution = 0.16 cm$^{-1}$, and imaging resolution = 702 nm for the EG and GIC samples.

**3.5 X-ray Diffraction (XRD):** A PANalytical Empyrean Diffractometer (Malvern Panalytical Ltd, UK) produced the information regarding crystal structure of EG compared to GIC using Bragg-Brentano focusing geometry at room temperature. 3kW Cu Kα radiation ($\lambda$ = 1.5406 Å) with a scan range of 2θ = 5 to 80° and step size of 0.013°.

**4. Result & Discussion**

**4.1 Microstructures and morphologies of fillers & polymer composite through FE-ESEM:**

FE-ESEM plays a crucial role in detail investigation of the fillers (EG & GNP) and composites (EG/PEID & GNP/PEID) morphology. The GIC expands through the thickness direction of GIC and expands upto highest expansion ratio of 280. GIC, we have used, is made from +50mesh, the through thickness structure is presented here Fig 1. After the expansion process of intercalated graphite compound turns into worm like structure of EG as shown in Fig 3a. Montage area mapping has been used to measure the average length and diameter of the EG filler achievable even after the expansion. At magnification of 20X, the average length and diameter are found to be 6-8mm and ~250μm. Due to intercalation, some packed layers are loosely connected, thus creates multilevel structural hierarchy as shown in Fig 3b with red circles. The puffed up, worm shaped material exhibits beneficial interconnection between the graphene sheets through the length. At higher magnification shown in Fig 3c, the diamond shaped pores (indicated by pink color) are visible throughout the single EG filler and pores are measured to be nm to ~20-30micron scale. Interestingly Fig 3d represents the flower alike structure of EG filler where graphite layers are clamped on one edge, the layers make interconnected branches, and on the

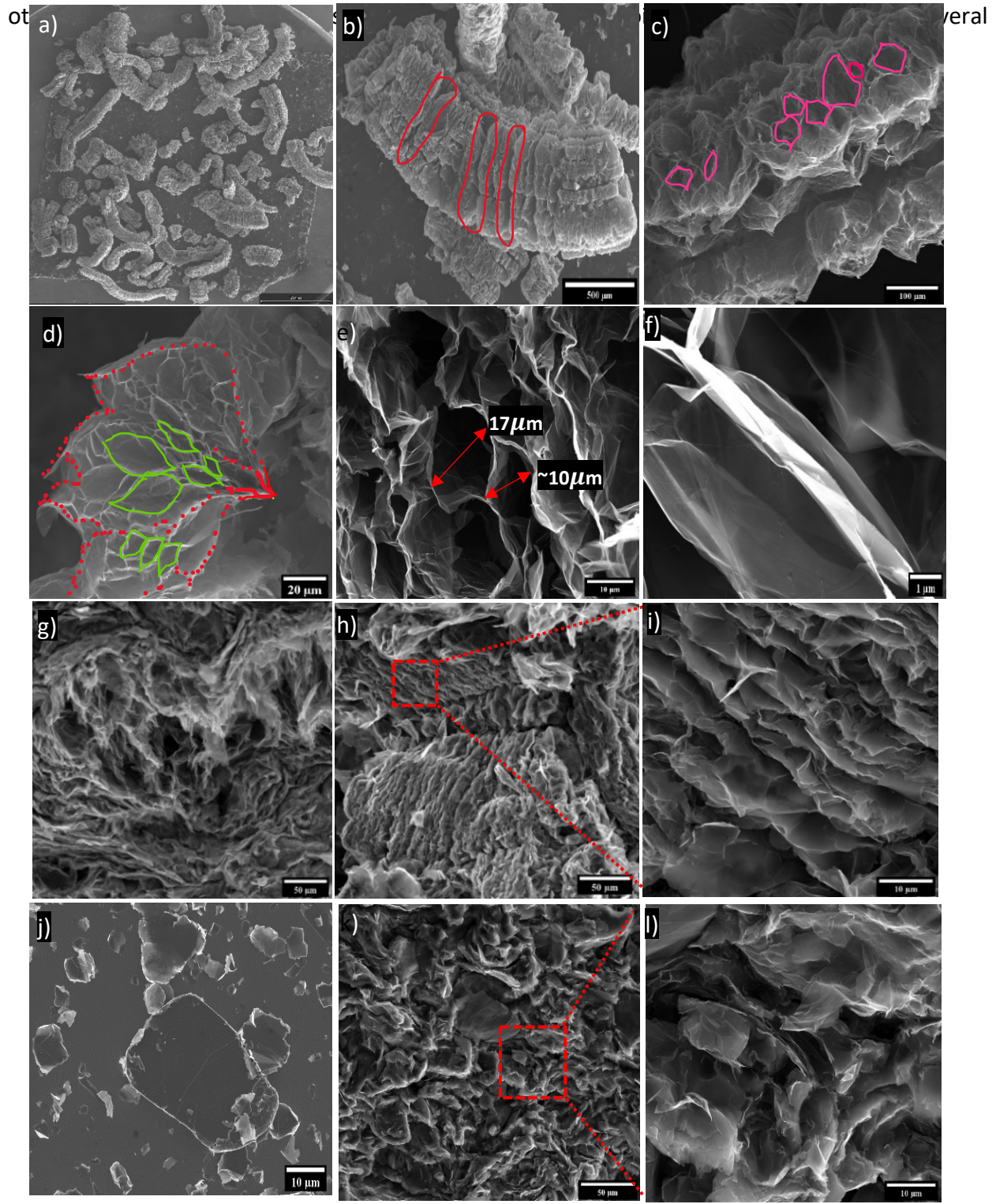

Figure 3 a)Montage area mapping of EG filler, FE-ESEM image of EG filler at b)30X c)150X d)500X e) 1500X & f)8000X. FE-ESEM images of EG/PEID composite for g) 5wt% at 350X, (h & i) 10wt% at 350X & 3500X. FE-ESEM images of j) GNP, (k & l) GNP/PEID composite for 10wt% GNP filler at 350x and 1500X layers to shape the flowers. Fig 3e shows the magnified view of EG pores of different length (10-20 micron) and wall thickness (~10nm), creating a continuous 3D network of graphene sheets within single expanded graphite. Such formation of network within the EG filler impairs higher electrical and thermal conductivity of composite after mixing with polymer. Thickness of graphene sheet is found to be ~10-20nm as shown in magnified view of Fig 3f, which is significantly lower than the length of the graphene nanosheet in EG or average diameter of EG filler, retaining a high aspect ratio.

FE-ESEM analysis of EG/PEID & GNP/PEID composites has been performed to examine the morphology of the fractured surface of solution casted nanocomposites. FE-ESEM images of 5 and 10 wt% EG filler composition with PEID are presented in Fig 3g & h at magnification of 350X and Fig 3i at magnification of 3500X. Even at lower mag, the porous structure is visible and graphitic network is distributed over the composites. Fig 3i displays the higher magnification SEM images of 10wt% EG/PEID composites, where the connected graphitic layers are clearly visible, occupied with PEID polymer through solution casting technique. The pores surrounded by graphene sheets inside EG adsorbs the liquid form of PEID polymer during stirring and curing process of composite preparation at higher temperature. Initial absorbance of DMAC solvent in EG makes the dispersion of EG filler into polymer solution easier by wetting the porous structure. Furthermore, agitation in mixture of liquid PEID and EG filler in DMAC solvent by ultrasonication prompts better diffusion of polymer into EG pores. Also, two minutes ultra-sonication at 20%

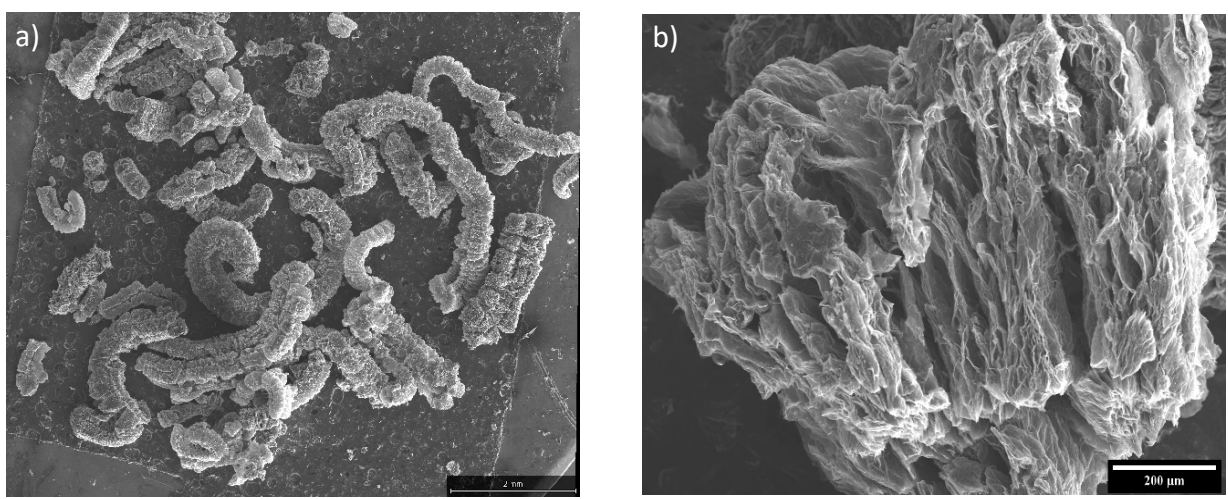

Figure 4 a) Montage area mapping of EG filler and b) FE-ESEM image of single EG filler after 2min ultrasonication at 20% amplitude in DMAC solvent amplitude of filler into polymer was optimized to achieve the quality dispersion for composite. Montage area mapping confirms that 2min sonication of the EG filler in DMAC solvent preserves the porous structure (as shown in Fig 4a & b) at large extent. Also, DMAC has lower viscosity than PEID solution, as a result the ultrasonication effect benefits more in liquid polymer, preserving the interconnected 3D network in composite. Such short time sonication also imparts partial exfoliation of EG into thinner & smaller graphene/graphite particles. Unique advantage of large sized EG and insignificant smaller particles constructs better conduction network through this solution mixing technique and reveals promising structural morphology of EG/polymer composite to obtain superior electrical & thermal properties. To compare, morphology of GNP filler and GNP/PEID composite is also presented in Fig 3j and 3k-l respectively. Average diameter of GNP filler is measured to ~25μm, significantly lower than EG filler. Fig 3k &l represents blended EG polymer composite image where polymer surrounds graphene, unable to establish any continuous network thus, raising high interfacial resistance between polymer and graphene, leading to lower electrical and thermal conductivity.

**4.2 XPS Analysis:**

**Table 1 Atomic composition by XPS analysis of GNP, EG and GIC**

|  | Atomic Composition by XPS (at%) | | |
| --- | --- | --- | --- |
|  | C (~285eV) | O (~532 eV) | S (~169eV) |
| GIC | 87.85 | 10.2 | 1.95 |
| EG | 91.85 | 6.91 | 1.24 |
| GNP | 97.48 | 2.52 | - |

XPS analysis was further performed to investigate the concentration of carbon (C), oxygen (O) and sulfur (S) elements before and after the thermal expansion of GIC as presented in table 3. Low sulfur content EG is desirable for less harmful composite preparation technique as well as improved properties of EG/polymer composite. Fig. 5a shows that two peaks of C1s and O1s at ~285eV and ~532eV are present for GIC and EG filler. S2p peak at ~169eV is visible in GIC as well as in EG filler spectra because of the included intercalated compounds. Due to the expansion process the sulfur content decreases from 1.95at% (in GIC) to 1.24 at% (in EG). Atomic percentage of carbon increases from 87.85% to 91.85% and oxygen reduces from 10.2% to 6.91% after expansion. Oxygen content decreases due to the thermal expansion at 900°C and increased atomic percentage of carbon content indicates the presence of more graphitic content in EG. XPS spectra of GNP has been also presented in Fig 5a and table 1, indicating the presence of 97.48at% C1s and 2.52at% O1s. The presence of increased oxygen groups in EG allows favorable interactions with oxygen groups in PEID through hydrogen bonding, leading to efficient interfacial thermal

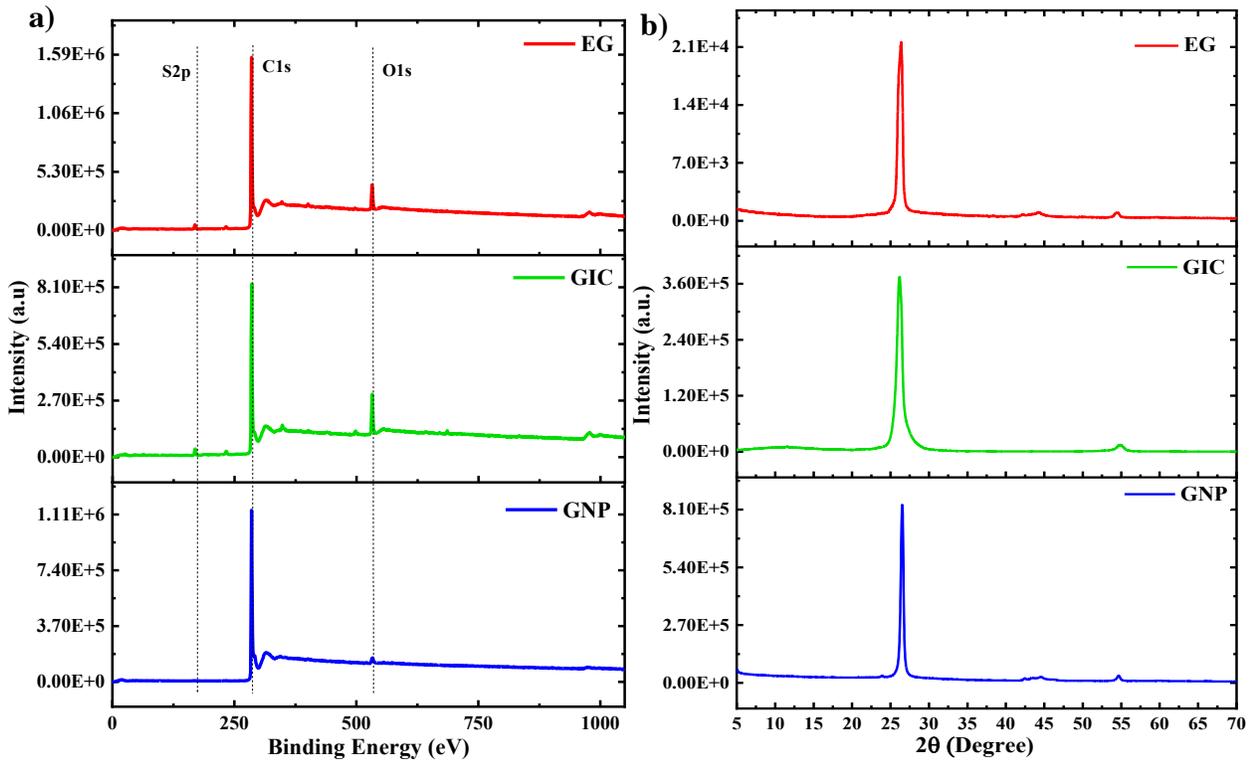

Figure 5 a) XPS spectra and b) XRD analysis of GNP, GIC and EG filler.

transport. This further enhances the conductivity of EG/PEID composite.

## 4.3 XRD analysis:

X-ray diffraction (XRD) analysis was performed to determine the crystal structure and interlayer spacing of GIC, EG and GNP filler. Fig. 5b shows a strong diffraction peak at $2\theta= 26.2°$ (002) for GIC, slightly shifted from the case of natural graphene case $2\theta= 26.5°$[57-60], implying highly oriented carbon material. The small shift in peak for GIC is attributed to the presence of intercalated compounds. From Bragg's diffraction equation, $n\lambda = 2d \sin \theta$, the interlayer spacing, d changes from 0.335nm of pristine graphite to 0.339nm for GIC. A broad, weak peak at $2\theta=11.2°$ (001) in GIC spectra indicates the attachment of functional groups through intercalation process in between graphene layers. The small shift in peak for GIC is attributed to the presence of intercalated compounds. On the contrary, a reduced sharp peak is visible at $2\theta= 26.35°$ (002)[61] for EG, still presenting interlayer spacing of 0.337nm, more closer to the graphitic carbon structure. A very weaker peak (0 0 2) is observed for EG case compared to GIC which is caused due to disorder in graphitic morphology [62] after expansion process. Still, a mostly aligned peak position in EG indicates the existence of intact chemical structure of graphite and interlayer order[63, 64]. This interconnected and stacked structure of EG enables better thermal transport throughout the polymer composite[65].

## 4.4 Raman Analysis:

Raman spectroscopy has been performed to further investigate the crystal structure before and after the thermal expansion. Particularly, identifying defective state of carbon structure using Raman spectroscopy has been widely used as nondestructive analysis. Raman spectroscopy of Fig. 5b exhibits three characteristic peaks of G-band, D-band and 2D-band at ~1580 cm$^{-1}$, ~1350 cm$^{-1}$ and ~2700 cm$^{-1}$ for graphite[66]. The G band denotes the in-plane vibration of the carbon-carbon bonds, results from stretching of defect-free sp$^2$ carbon of hexagonal ring [67] and the D band signifies the vibrational mode, caused due to the amorphous disordered structure of sp$^3$ hybridized carbon[68]. 2D band also can be seen at around 2700 cm$^{-1}$ and represents a second-order two phonon process[69]. Raman spectra of GIC and EG clearly shows the representative peaks of G, D & 2D bands. Broad D band in GIC is present and cases $I_D/I_G$ ratio of 0.21, suggesting the product of structural defects due to intercalation process. On the contrary, D band is negligible in EG Raman spectra, refers to highly ordered defect-free graphite structure. The absence of a D band in EG suggests the presence of a highly ordered defect-free graphite structure in EG. The peak shape of the 2D band varies with the number of graphene layers. The 2D peak position and shape of GIC and EG filler shows slight change after thermal treatment, indicating the reduction in van der Waals interaction between graphene layers and the increasing of the interlayer distance between these layers[70]. Because of this change in graphitic structural order, EG imparts strong electrical and thermal properties in EG/PEID composite through the graphitic nanosheets.

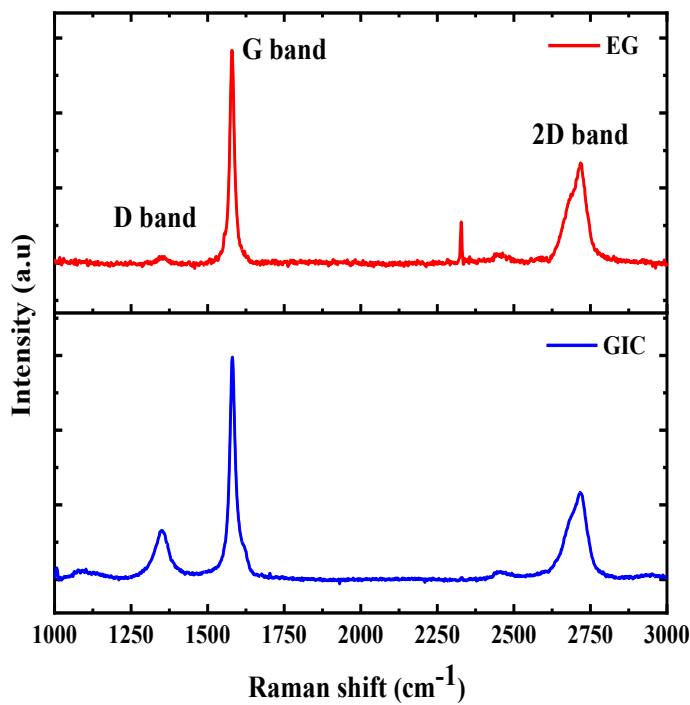

Figure 6 Raman spectra of GIC and EG filler.

### 4.5 Electrical Properties

Figure 7a and b represent electrical conductivity and resistivity of the EG/PEID composites as a function of EG filler content. Comparison in electrical conductivity and resistivity with GNP/PEID composite is also presented as a function of GNP filler content in Fig 7a & b. Electrical conductivity is an intrinsic property of polymer composite, has been investigated in this work. The conductivity of pure PEID is about $1.2 \times 10^{-17}$ Sm$^{-1}$ [51, 71] has been beneficial property for electronic packaging material application. The value of EC starts to increase with addition of EG filler. Starting at 1wt% EG filler amount, EC of pure polymer rises suddenly from $1.2 \times 10^{-17}$ Sm$^{-1}$ to $8.3 \times 10^{-3}$ Sm$^{-1}$. In contrast the GNP/PEID composite reaches at EC value of $7.1 \times 10^{-3}$ Sm$^{-1}$ for 2.5wt%. EG filler at only 1wt% already starts to touch each other establishing a percolative network due to the continuous conductive network in the composite. Increasing EG in the PEID matrix, causes a sharp increase in conductivity at 1wt% EG filler and keep increasing till 2wt%, reaching at ~17.8 Sm$^{-1}$. It is well known that the filler concentration at and above which drastic increase in conductivity occurs due to tunneling effect is called the percolation threshold, found at 1 wt% in this study. From 2.5wt% EG filler inclusion, the EC value of EG/PEID composite increases at slower rate upto 10wt% filler concentration. The conductivity of the 10wt% EG/PEID composite reaches at 969 Sm$^{-1}$, showing ~10$^{19}$ fold higher EC than of pure PEID. And 10wt% GNP/PEID composite shows enhancement of 2.86 Sm$^{-1}$, indicating ~ 100% decrease compared to PEID/EG composite. Such higher enhancement of EG/PEID composite is subjected

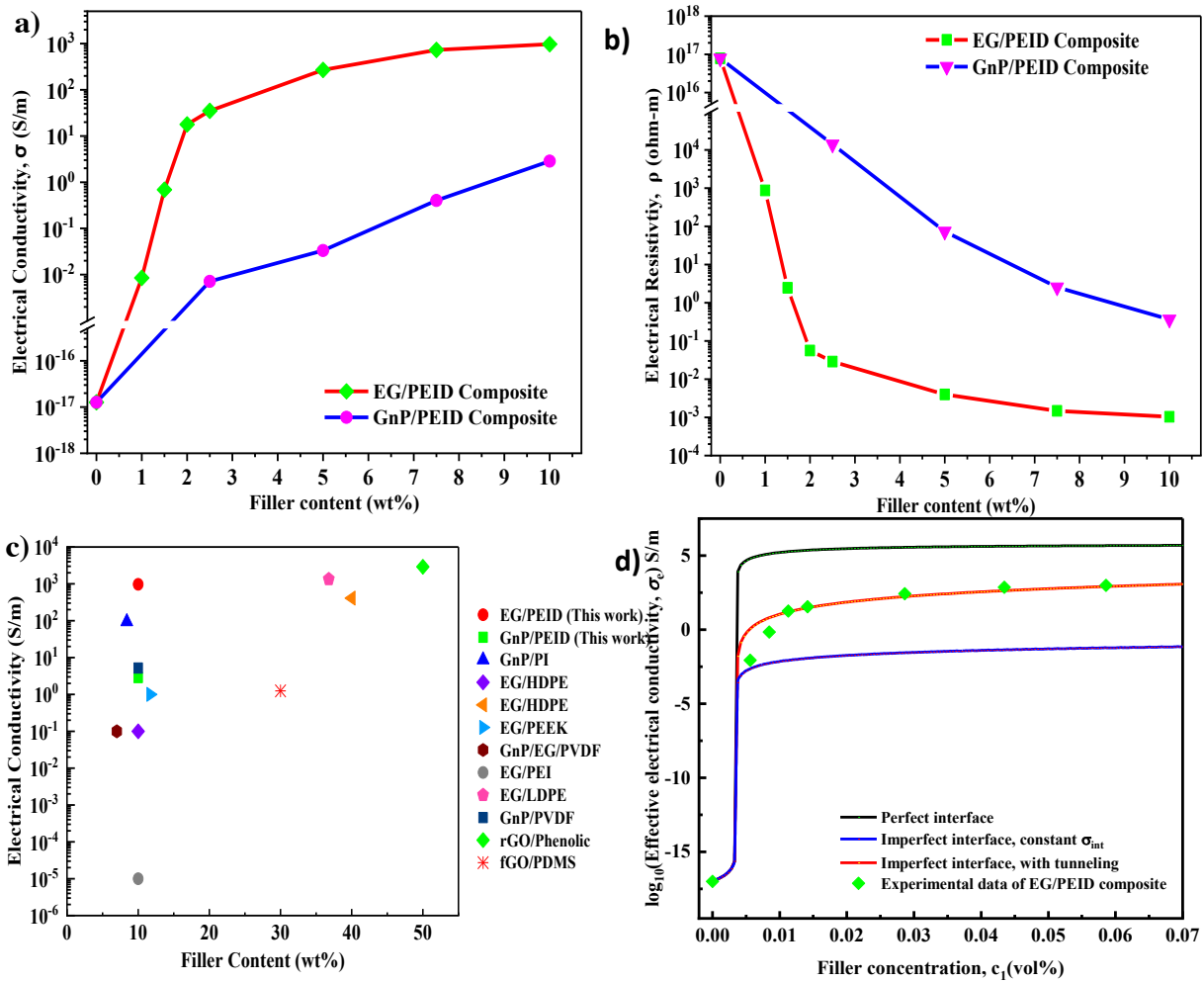

Figure 7 a) Electrical conductivity value of the PPS/EG nanocomposites as a function of EG, b) electrical resistivity of the PPS/EG nanocomposites as a function of EG, c) comparison of electrical conductivity value of polymer composites with previous works, and d) The effective in-plane electrical conductivity in EG/PEID composites with respect to EG filler volume concentrations under perfect and imperfect interface conditions to beneficial porosity of EG filler itself. Morphology of EG filler in the composite has been understood in detail through the FE-SEM images. Solution casting technique followed by short time sonication helps to disperse the filler, also filling the liquid polymer into the porous EG. Simultaneously, DMAC solvent makes the EG filler's porous structure wet enough to diffuse the polymer into it, thus makes the filler to polymer conduction path interconnected and efficient electrical conduction takes place. Fig 7b supports the influence of EG filler in lowering the resistivity of polymer through the addition of EG filler. As resistivity is inversely proportional to the EC value, it represents that dramatic reduced value due to inclusion of EG filler, reaching at $1.4 \times 10^{-3}$ $\Omega$m for 10wt% composition whereas similar composition of GNP/PEID composite, the value reaches only at $3.6 \times 10^{-3}$ $\Omega$m. Different EC value for composite with different weight percentage of EG or graphene filler and comparison with different preparation technique are presented in Fig 7c and table 2 respectively, indicating superior EC value achieved in this study at lower content of filler.

**Table 2 Comparison of EC for different polymer-graphene and EG-polymer composite**

| Filler | Matrix | Fraction | ρ (S/m) | Preparation method | ref |
|---|---|---|---|---|---|
| GnP | PI | 5 vol% | 94 | solvent casting using N-methyl-2-pyrrolidone (NMP) | 1 |
| EG | PEI | 10wt% | 1E-5 | Ultrasonic melt-extrusion | 72 |
| EG | LDPE | 36.78wt% | 1345 | Melt mixing | 42 |
| EG | PDS | 10wt% | 2500 | Solvent casting | 73 |
| EG | HDPE | 10wt% | 0.1 | Melt mixing | 74 |
| EG | HDPE | 40wt% | 410 | Melt mixing | 75 |
| EG | PEEK | 11.7wt% (7vol%) | $2.40 \times 10^{-3}$ | Melt mixing | 76 |
| Graphite/EG | PVDF | 7/0.5wt%/wt% | 0.1 | Melt Crystallization and quenching in water | 77 |
| GnP | PVDF | 10wt% | 5.2 | Solvent blending–casting | 78 |
| GnP | PS | 30wt% | 1.25 | Solution mixing | 79 |
| rGO | Phenolic | 50 wt% | 2900 | Solution mixing | 80 |
| rGO | PDMS | 0.4wt% | 0.32 | Bidirectional freeze-casting | 81 |
| EG | PEI | 10wt% | 969 | Solution mixing | This work |
| GnP | PEI | 10wt% | 2.86 | | |

PI: Polyimide; PEEK: Poly(ether ether ketone); HDPE: High density polyethylene; LDPE: Low density polyethylene; PS: Polystyrene; PVDF: Polyvinylidene fluoride; PDMS: Polydimethylsiloxane

Practically, 3D interconnected network as well as the intersheet contact resistance crucially need to be considered to achieve effective EC of composite. Graphene nanofiller imparts insignificant through thickness EC because of weak van der Waals force in between graphene layers which makes it difficult to obtain desired electrical properties of composite. In EG/PEID composite, continuous graphitic network of 3D filler is present as shown in FE-ESEM images which allows to take the advantage of in plane EC value ($10^7 - 10^8$)[82] of graphene nanosheets, thus leading to higher EC value of composite. Simultaneously, 3D interconnected network reduces the percolation threshold of composite, hereby use of higher content of filler can be avoided As the large graphite nanosheets are covalently bonded even at the bend or turns of these worms like structure of EG even in the EG/PEID composite, that takes part in conduction between polymer and EG. This kind of structure also overcomes the intersheet resistance issue in between graphite nanosheets developing higher number of contacts. In addition, PEID inherently has oxygen and nitrogen groups that interacts with oxygen functional groups present in the EG (as shown in XPS analysis) through hydrogen bonding, leading to strong interaction between polymer and graphene nanosheets of the EG.

**4.6 Effective Medium Theory:**

The enhancement in EC of EG/PEID composite occurs due to several factors- (1) continuity of large graphite nanosheet over the length scale of EG; (2) efficiency in charge transfer between polymer and graphite sheets as the pores/gaps becomes occupied with polymer during solution casting technique; (3) percolative network at low concentration of filler. We have implemented here effective medium model to calculate in-plane electrical conductivities by varying interfacial resistance depending on interfacial layer and interfacial tunneling between the fillers and the matrix as presented in Fig 7d.

Xia et al. calculated the effective electrical conductivity tensor, $L_e$, of the two-phase medium considering polymer matrix as phase 0 and EG inclusions as phase 1 using following equation:

$$c_0[(L_0 - L_e)^{-1} + S_0 L_e^{-1}]^{-1} + c_1 \langle [(L_1 - L_e)^{-1} + S_1 L_e^{-1}]^{-1} \rangle^* (\theta) = 0 \qquad (1)$$

where $c_0$ and $c_1$ are the volume fractions of phase 0 and 1 respectively. $S_i$ and $L_i$ represent the Eshelby's S-tensor and electrical conductivity tensor of phase i and subscript 'e' refers to the 'effective' properties of the composites. The in-plane and out-of-plane effective electrical conductivities $\sigma_1^e$ and $\sigma_3^e$ can be obtained by solving the these following nonlinear equations:

$$c_0 \frac{\sigma_1^e(\sigma_0-\sigma_1^e)}{\sigma_1^e+S_{11}^0(\sigma_0-\sigma_1^e)} + c_1 \left\{ A(\theta) \frac{\sigma_1^e(\sigma_1-\sigma_1^e)}{\sigma_1^e+S_{11}(\sigma_1-\sigma_1^e)} + B(\theta) \frac{\sigma_3^e(\sigma_3-\sigma_3^e)}{\sigma_3^e+S_{33}(\sigma_3-\sigma_3^e)} \right\} = 0 \qquad (2)$$

$$c_0 \frac{\sigma_3^e(\sigma_0-\sigma_3^e)}{\sigma_3^e+S_{33}^0(\sigma_0-\sigma_3^e)} + c_1 \left\{ A(\theta) \frac{\sigma_1^e(\sigma_1-\sigma_1^e)}{\sigma_1^e+S_{11}(\sigma_1-\sigma_1^e)} + B(\theta) \frac{\sigma_3^e(\sigma_3-\sigma_3^e)}{\sigma_3^e+S_{33}(\sigma_3-\sigma_3^e)} \right\} = 0 \qquad (3)$$

where $\theta$ indicates orientational average in the angle, and the subscripts '1' and '3' refer to the components in in plane and out-of-plane directions, respectively. $\sigma_0$, $\sigma_1$ and $\sigma_3$ indicate the electrical conductivity of a polymer matrix, in-plane and out-of-plane of EG filler accordingly. Also, the $\theta$-dependent coefficients, $A(\theta)$, $B(\theta)$ and $C(\theta)$ are presented below:

$$A(\theta) = \frac{9+2\cos\theta+\cos 2\theta}{12}, B(\theta) = \frac{3-a\cos\theta-\cos 2\theta}{12}, C(\theta) = \frac{3+2\cos\theta+\cos 2\theta}{6} \qquad (4)$$

$S_{11}^0$ and $S_{33}^0$ are the two components of the S-tensor for the matrix phase, taking the statistical distribution of EG fillers to be isotropic, leading to $S_{11}^0 = 1/3$ and $S_{33}^0 = 1/3$. Two independent components $S_{11}$ and $S_{33}$ are the shape factors related to aspect ratio and can be calculated using following equations:

$$S_{11} = S_{22} = \frac{\alpha}{2(1-\alpha^2)^{3/2}}[\arccos\alpha - \alpha(1-\alpha^2)^{1/2}]; \alpha < 1$$

$$S_{33} = 1 - S_{11}$$

where, $\alpha$ is the aspect ratio of graphite nanosheet.

Two major sources of interfacial effects are mainly responsible for the reduction of electrical conductivity of polymer composite. The first one refers to the imperfect interface between the inclusions and the polymer matrix, which tends to lower the overall conductivity. Introducing a thin interfacial layer around the inclusions can be implemented to treat the interface effect. Overall electrical conductivity for imperfect interface can be obtained by assuming the interfacial layer as isotropic and the equations are presented below:

$$\sigma_i^{(C)} = \sigma^{(int)} \left[ 1 + \frac{(1-c_{int})(\sigma_1-\sigma^{(int)})}{c_{int}S_{11}(\sigma_1-\sigma^{(int)})} \right] \qquad (5)$$

Where $\sigma^{(int)}$ is the isotropic electrical conductivity of the interfacial layer, and $c_{int}$ is the volume fraction of the interfacial layer around the inclusions.

$$c_{int} = 1 - \frac{\lambda}{2}\left(\frac{\lambda}{2\alpha}\right)^2 / \left(\frac{\lambda}{2}+h\right)\left(\frac{\lambda}{2\alpha}+h\right)^2$$

And the second effect is the tunneling effect of electron on interfacial conductivity. According to Wang et al., the effect starts to develop with the increment of filler loading and decrease in distance between inclusive filler, causing greater probability of electron tunneling in the start-up zone of percolative network and formation of nanocapacitors at the interface. The tunneling-assisted interfacial conductivity can be evaluated using the equation below:

$$\sigma^{(int)} = \sigma_0^{(int)} / \tau(c_1, c_1^*(\theta), \gamma_\sigma) \quad (6)$$

where $\sigma_0^{(int)}$, $\tau$, $\gamma_\sigma$ and $c_1^*(\theta)$ represents the intrinsic electrical conductivity of the interfacial layer, resistance-like function, electron tunneling scale parameter, and $\theta$-dependent percolation threshold respectively. Derivation of this parameters has been further provided in the supplementary information.

**Table 3 Physical values used in the theoretical calculation of the effective conductivity of nanocomposites.**

| Material Parameters | Values |
|---|---|
| Aspect ratio of the graphene filler, $\alpha$ | $1.00 \times 10^{-3}$ |
| Graphene half thickness, $\lambda$ | $8.00 \times 10^{-9}$ |
| Thickness of the interlayer, h (m) | $10.00 \times 10^{-9}$ |
| In-plane electrical conductivity of EG, $\sigma_1$ (S/m) | $8.32 \times 10^7$ |
| Out-of-plane electrical conductivity of EG, $\sigma_3$ (S/m) | $8.32 \times 10^2$ |
| Electrical conductivity of the PEID polymer, $\sigma_0$ (S/m) | $1.2 \times 10^{-17}$ |
| Electrical conductivity of the interface at $c_1$=0, $\sigma_0^{(int)}$ (S/m) | $9.5 \times 10^{-3}$ |
| Scale parameter of the electronic tunneling at the interface, $\gamma_\sigma$ | $1.3 \times 10^{-5}$ |

Electrical conductivity for EG/PEID composite has been explored with this effective medium model to observe the effect of imperfect interface with interfacial layer and electron tunneling effect. Effective electrical conductivity of composite for three different interface conditions is plotted for randomly oriented EG filler into polymer matrix by assuming $\theta = \pi/2$ in Fig 7d. Table 3 presents all the value of the parameters, used to calculate effective electrical conductivity. A perfect interface between the EG filler and polymer matrix possess the result (shown by black curve) with highest EC value assuming no interfacial resistance between filler and polymer by solving equation 3 & 4. Induced imperfect interface with a constant $\sigma_0^{(int)}$ causes interfacial resistance, attributing a significant drop in the effective conductivity (as shown by blue line), which is directly related to the interfacial resistance between polymer and EG filler. The calculated EC value for imperfect interface reveals that the contribution of electron tunneling effect adds a positive impact as indicated by red curve. For all composition of EG/PEID composite, the fillers tend to preserve the contact between graphite layers, that reduces the interfacial resistance and increases the probability of this tunneling effect. A conductivity due to this tunneling effect is then added to electrical conductivity with imperfect interface. Fig 7d shows that the theoretical approach of imperfect interface with tunneling effect is in good agreement with experimental value of EG/PEID composite.

### 4.7 Thermal Conductivity data:

Thermal conductivity of composite is a fundamental property to understand how the heat conduction rate got improved via the addition of carbon-based filler. Solution casting technique has been utilized to prepare the composite separately based on two different carbon-based fillers (EG & GNP) and through thickness thermal conductivity has been measured to compare their contribution in thermal properties. Fig 8a represents the measured through thickness thermal conductivities of EG/PEID and GNP/PEID composites as a function filler concentration of EG & GNP and $k$ value has been improved with the increasing filler content.. $k$ of pure PEID has been measured in this work and found to be 0.23 W/mK, matches with literature review. Inclusion of only 2.5wt% EG filler raises $k_{EG/PEID\ composite}$ value to 3.4 W/mK, representing 1378% enhancement with respect to pure PEID. On the contrary, $k_{GNP/PEID\ composite}$ (0.58 W/mK) only shows an enhancement of 152% for 2.5wt% GNP filler. Addition of 10wt% EG filler enhances the composite $k$ value to 7.3 W/mK, indicating ~3100% enhancement in $k$ value compared to pure polymer, whereas $k$ of GNP/PEID composite has been enhanced by 947% for similar content of GNP filler concentration. Such significant enhancement in EG/PEID composite can be referred to the reason of 3D interconnected continuous network of EG filler, as well as partially exfoliated graphite nanosheets, creating efficient heat transfer pathway in the composite. Intrinsic high thermal conductivity of > 300 Wm$^{-1}$K$^{-1}$ of EG imparts its excellent impact on the composite $k$ value.

In addition, solution casting technique using DMAC solvent apparently allows the liquid PEID polymer to get diffused into the pores of the 3D EG filler as shown by FE-ESEM images and remains beneficial in avoiding further processing steps to prepare the composites. The enriched number of pores in EG structure adds their unique advantages on composite's ultimate properties.

However, preserving this porous structure of EG filler and lateral size of graphite nanosheets in EG (~250 μm) is a key challenge during the composite preparation process. Exerting minimal force is immensely required in maintaining the microstructure of EG as well as the lateral size in composite at low content of filler. Solution casting technique has an utmost influence in preserving the structure in composite and enhancing the $k_{EG/PEID\ composite}$ to such large extent.

On the other hand, short time cavitation force using ultrasonication further can prompt the adsorption of polymer into nanoscale sized pores of EG fillers. We investigated the effect of the ultrasonication on the $k$ value of composite by preparing the 10wt% EG/polymer composite at different ultrasonication time. The trend in measured $k_{EG/PEID\ composite}$ value reveals that ultra-sonication time of EG filler into polymer has direct influence on thermal conductivity value as

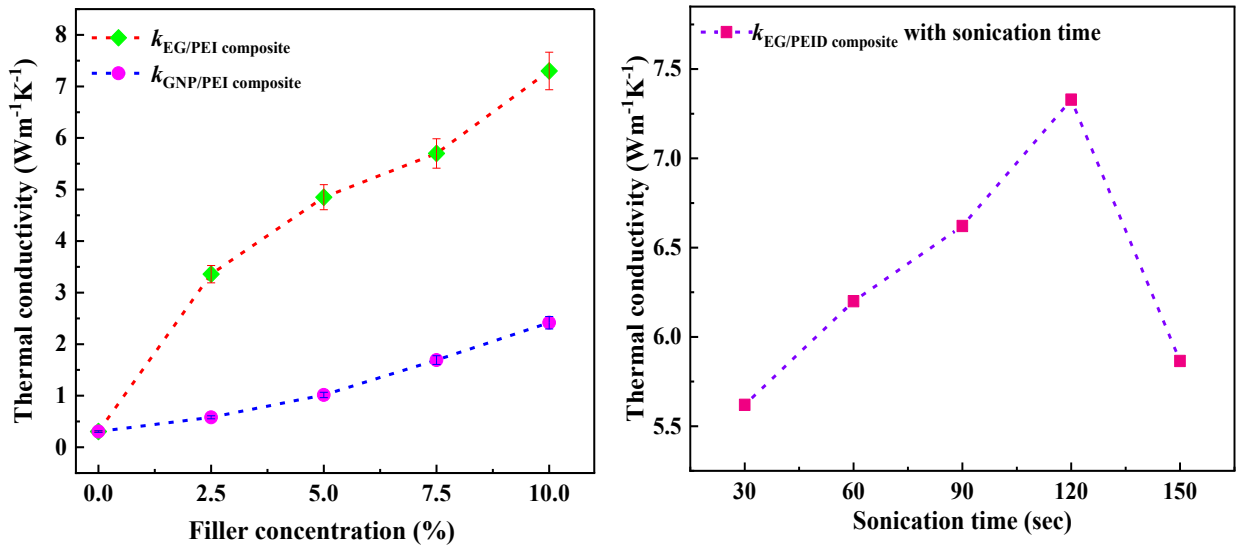

Figure 8 a) Thermal conductivity value of EG/PEID & GnP/PEID composite as a function of different filler content, b) $k$ value of EG/PEID composite with different sonication time at 20% ultrasonication power shown in Fig 8b. The increased sonication time from 30sec to 120sec leads to value of $k_{EG/PEID\ composite}$ from 5.6 W/mK to 7.3W/mK, then it reduces the $k$ value to 5.W/mK for 150sec. Accordingly, the vibration time by ultrasonication has been optimized to obtain the highest $k_{EG/PEID\ composite}$ value.

Lower $k_{GNP/PEID\ composite}$ implies the higher interfacial resistance between polymer and graphene due to the obstacle induced by polymer between graphene-graphene contact in heat conduction, which ultimately requires high content of filler to achieve higher k. Kargar et al.[83] reported k value of graphene epoxy composite ≈ 11W/mK at very high graphene filler content of 45vol%. Functionalization via oxidation has been chosen as a promising way to enhance the interfacial thermal conductance in several studies[ref]. Presence of oxygen functional groups (presented in XPS analysis) on EG filler due to intercalation process can be referred to the reason of reduced interfacial thermal resistance between EG and polymer. Overall, the obtained $k_{EG/PEID\ composite}$ value illustrates the fact of continuous conductive network of EG has been perseverd in composite via the solution casting technique followed by short time sonication, which outperforms the $k_{GNP/PEID\ composite}$ value.

## 5. Conclusion:

In summary, a high electrically and thermally conductive EG/PEID composite has been developed using a solution casting technique followed by short time ultrasonication at a low content of 10wt% EG filler. 10wt% filler content EG/PEID composites displays a superior electrical conductivity of 969Sm$^{-1}$ with percolation threshold at significantly lower EG filler concentration of 1.5wt%. We have implemented effective medium approach to observe the effect of electron tunneling resistance on electrical conductivity of EG/polymer composite. A dramatic enhancement of ~3074% has been obtained for EG/PEID composite k value of 7.3W/mK, with respect to pure polymer, which is also 3 times higher than GNP/PEID composite k value at 10wt% inclusion of filler. FE-ESEM images provide evidence with the detailed morphology of porous EG filler and EG/PEID composite. It is found that 3D continuous network of EG filler is responsible for achieving superior performance in electrical and thermal properties. This strategy provides an efficient way to occupy the porous structure of EG filler with polymer, allowing the formation of well-connected conductive pathway to develop high electrical and thermal conducting composites. This polymeric material shows enormous potential in wide range of thermos-electronic application.

**Acknowledgment**

FT, SD, JG and AN acknowledge support from National Science Foundation CAREER award under Award No. #1847129. We also thanks Mohammed Ibrahim, PhD, from Thermo Fisher Scientific for collecting the Raman spectra.

**Conflicts of interest**

There are no conflicts to declare.